**On-Command Disassembly of Microrobotic Superstructures for Transport and Delivery of Magnetic Micromachines**

Fabian C. Landers†, Valentin Gantenbein†, Lukas Hertle, Andrea Veciana, Joaquin Llacer-Wintle, Xiang-Zhong Chen, Hao Ye*, Carlos Franco, Josep Puigmartí-Luis, Minsoo Kim*, Bradley J. Nelson, Salvador Pané*

* Corresponding authors.
† These authors contributed equally to this work.

F. C. Landers, V. Gantenbein, L. Hertle, A. Veciana, J. Llacer-Wintle, H. Ye, C. Franco, M. Kim, B. J. Nelson, S. Pané
Multi-Scale Robotics Lab, Institute of Robotics and Intelligent Systems, ETH Zurich; Tannenstrasse 3, 8092 Zurich, Switzerland
E-mail: haoyeh@ethz.ch, minkim@ethz.ch, vidalp@ethz.ch

X.-Z. Chen
Institute of Optoelectronics, Shanghai Frontiers Science Research Base of Intelligent Optoelectronics and Perception, Fudan University, Shanghai 200438, People's Republic of China
Yiwu Research Institute of Fudan University, Yiwu 322000, Zhejiang, People's Republic of China

J. Puigmartí-Luis
Departament de Ciència dels Materials i Química Física, Institut de Química Teòrica i Computacional, University of Barcelona; Martí i Franquès, 1, 08028, Barcelona, Spain
Institució Catalana de Recerca i Estudis Avançats (ICREA); Pg. Lluís Companys 23, Barcelona 08010, Spain

**Keywords:** Microrobots, Magnetic navigation, Magnetic hyperthermia, Targeted delivery, Two-photon lithography

**Abstract**

Magnetic microrobots have been developed for navigating microscale environments by means of remote magnetic fields. However, limited propulsion speeds at small scales remain an issue in the maneuverability of these devices as magnetic force and torque are proportional to their magnetic volume. Here, we propose a microrobotic superstructure, which, as analogous to a supramolecular system, consists of two or more microrobotic units that are interconnected and organized through a physical (transient) component (a polymeric frame or a thread). Our superstructures consist of microfabricated magnetic helical micromachines interlocked by a magnetic gelatin nanocomposite containing iron oxide nanoparticles (IONPs). While the microhelices enable the motion of the superstructure, the IONPs serve as heating transducers for dissolving the gelatin chassis via magnetic hyperthermia. In a practical demonstration, we showcase the superstructure's motion with a gradient magnetic field in a large channel, the disassembly of the superstructure and release of the helical micromachines by a high-frequency alternating magnetic field, and the corkscrew locomotion of the released helices through a small channel via a rotating magnetic field. This adaptable microrobotic superstructure reacts to different magnetic inputs, which could be used to perform complex delivery procedures within intricate regions of the human body.

## 1. Introduction

Over the last two decades, significant milestones have been accomplished in the field of biomedical magnetic small-scale robots.[1–3] Researchers initially focused on exploring basic magnetic micro- and nanoarchitectures, including structures like nanowires, helices, and microparticles.[4] Their primary goal was to understand the fundamental principles governing motion at the micro- and nanoscale and to engineer strategies for achieving controlled movement using external magnetic fields. Building on these initial investigations and leveraging advancements in magnetic navigation, material science, and manufacturing, small-scale roboticists are now pioneering the development of sophisticated small-scale magnetic machinery. These systems incorporate integrated functional building blocks,[5] such as containers for carrying therapeutic agents or cells,[6–10] stimuli-responsive materials for tissue ablation or cell stimulation,[11,12] contrast agents,[13] and/or sensing elements.[14,15]

Despite the undeniable progress in the field, substantial barriers lie ahead in the journey to transitioning these small-scale robots from laboratory settings to the operating theatre. Challenges include the development of magnetic navigation systems compatible with healthcare facilities and imaging equipment,[16,17] as well as issues concerning the implantation procedure,[18] tracking and monitoring of these devices,[19] the therapeutic dosing for a specific condition or disease,[20] or biocompatibility and biodegradability features.[21] A primary issue concerns the maneuverability of these devices once these are implanted in the vascular system. As magnetic forces and torques are directly proportional to the magnetic moment of the device, small-scale robots (<100 μm) suffer from limited propulsion speeds, particularly in ultra-low Reynolds number regimes, making them unsuitable for practical biomedical applications.[22–24] Conversely, larger designs (100-1000 μm) can travel faster but cannot navigate through confined biological conduits. To address these issues, some solutions have been proposed. For example, Huang and co-workers have proposed a magnetic microtransporter comprising an Archimedes screw pumping mechanism capable of transporting and delivering magnetic micro particles and microhelices.[25] However, this system relies on a nondegradable large structure, thus requiring retrieval after the magnetic agents have been delivered. A recent strategy proposes the use of swarms of magnetic micro- or nanoparticles, which potentially can swim cooperatively in fluids thanks to their magnetic dipole-dipole interactions.[26,27] Interestingly, the morphology and locomotion mechanism of the swarms can be adjusted from ribbons to clusters as a function the applied magnetic inputs. A main drawback from these assemblies is that forces in the fluid vasculature of large animals or humans will likely disrupt and disaggregate the assembly, which could be disadvantageous in the context of medical procedures.

Here, we propose microrobotic hierarchical superstructures consisting of magnetic helical micromachines interlocked with a thermally responsive transient magnetopolymer chassis. By combining iron oxide nanoparticles (IONPs), possessing either polyacrylic acid (PAA) or polyethylene glycol (PEG) as ligands, we enabled the disassembly and release of the helical microcomponents on-command by applying high-frequency magnetic stimulation, which triggers the dissolution of the chassis by hyperthermia. The helical devices are made of metallic iron having superior magnetic properties, as this metal exhibits the highest saturation magnetization among magnetic materials with biocompatibility features.[28] To fabricate these architectures, we capitalize on our previous assembly-free manufacturing method that combines metal electrodeposition and polymer casting in 3D templates obtained by 3D direct laser writing.[29] This method allows for building 3D complex mechanically interlocked metal-organic microrobots. The proposed smart superstructures could be used to navigate and deliver small magnetic helical micromachines to intricate small vessels and capillaries, thereby allowing them to access difficult-to-reach anatomical sites within the human body.

## 2. Results and discussion

### 2.1. Design and fabrication of microrobotic superstructures

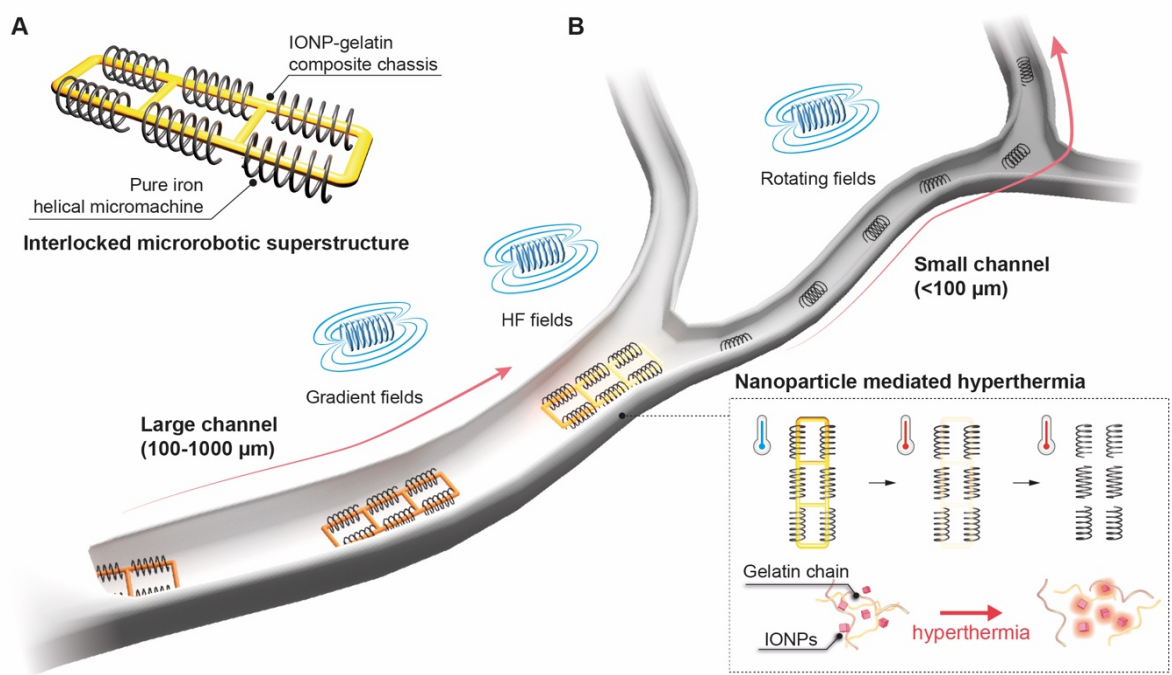

**Figure 1.** Schematic diagram of targeted delivery using microrobotic superstructures. (A) The prototype of the microrobotic superstructure is composed of an IONP/gelatin composite chassis, which carries pure iron helical micromachines. (B) In a larger channel, a gradient magnetic field propels the assembled chassis. Upon reaching a small channel, the helical micromachines are disassembled on-command by dissolving the chassis through the application of magnetic hyperthermia. An alternating high-frequency magnetic field (HF field) induces magnetic hyperthermia by the IONPs. Subsequently, the disassembled helical micromachines continue their movement within the smaller channel, exhibiting a corkscrew motion induced by a rotating magnetic field.

**Figure 1** schematically illustrates the concept of targeted delivery using a microrobotic superstructure prototype, navigating through a branched vessel system using external magnetic fields and magnetic gradients. The superstructure comprises a polymeric chassis made of an IONP/gelatin composite, which serves as structural scaffold. This chassis carries a collective of ferromagnetic iron helical micromachines, which enable successful steering of the superstructure through a large diameter vessel with magnetic field gradients at relatively high speeds. At a bifurcation at which the vessel is branching into smaller scale conduits, the superstructure cannot further advance due to its size. Then, the chassis can be dissolved via magnetic hyperthermia of IONPs, releasing its interlocked microhelices. These helices are capable of propelling through smaller branches by corkscrew locomotion, using rotating magnetic fields.[30]

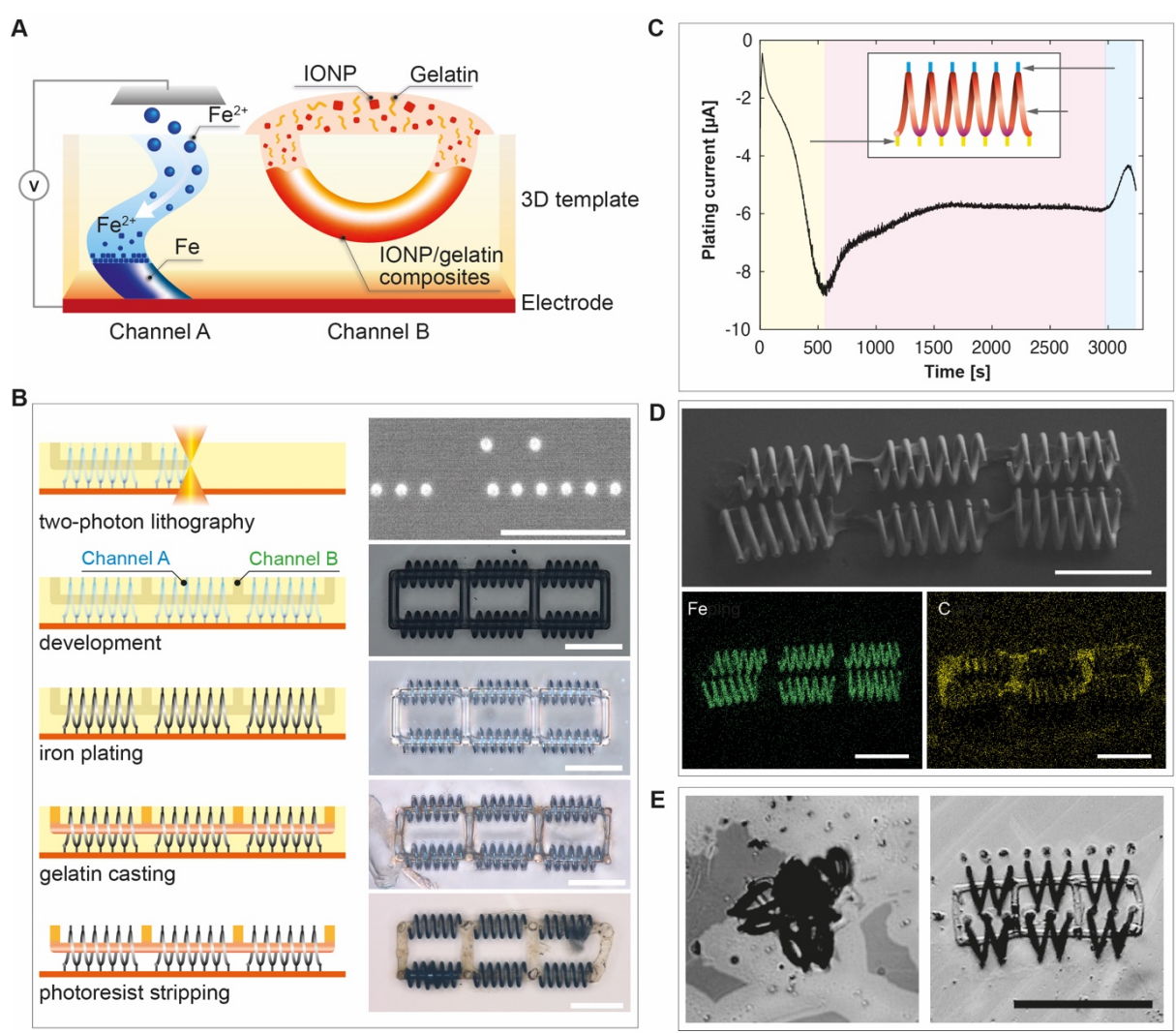

**Figure 2.** Fabrication of interlocked microrobotic superstructures. (A) Schematic of template-based microscale 3D architecture manufacturing process using heterogeneous materials (metal

through channel A and polymer through channel B). (B) Process overview with optical microscope images at each step. The scale bars indicate 75 μm. (C) Change of plating current over time. Significant fluctuations are observed in the plating current as the plating domain changes (yellow, red, and blue in sequence). It is used for monitoring the plating process. (D) Scanning electron microscope (SEM) and energy-dispersive X-ray spectroscopy (EDX) images of the interlocked microrobotic superstructure. The scale bars indicate 75 μm. (E) Comparison of pure gelatin chassis made of 5% (left) and 10% (right) of gelatin concentration. The scale bar indicates 100 μm.

The manufacturing of the interlocked microrobotic superstructures is schematically depicted in **Figure 2**A and B.[29] First, a conductive copper-coated glass substrate was covered with a photoresist layer. Direct laser writing, also known as two-photon lithography, was used to create a 3D template within the photoresist layer with two types of microchannels that are interweaved (Figure 2A). Microchannels type A, which are used for iron filling (i.e., helical micromachines), penetrate through the entire photoresist layer to establish electrical connectivity between the iron source (electrolyte) and the conductive substrate. The microchannels type B, made for gelatin casting (i.e., the chassis), are designed to avoid any contact with the conductive substrate. We conducted a parameter sweep (laser power and scan speed) to identify the ideal exposure parameters for writing the 3D channels, as detailed in Fig. S1 and Table S1 (Supporting Information). Underexposure resulted in inadequate electrical connections and inconsistent filling during subsequent electroplating and casting steps. Overexposure, conversely, could lead to the coalescence of the channels and subsequent uncontrolled electroplating into adjacent channels. Both the microrobot design and printing parameters were collectively optimized to establish the ideal design and fabrication conditions, as shown in Fig. S2 and S3 (Supporting Information). Because of the tendency of the photoresist to crack under the stress induced by subsequent heating and cooling cycles, the spacing between the microrobots was increased and the temperature was gradually ramped (Fig. S4, Supporting Information). By fine tuning the microchannel design, the electroplating process allowed to only fill the microchannels A with iron. During this process, we monitored the plating current to identify the growth of the plated microstructures, as evidenced in Figure 2C. As the current increases proportionally to the area parallel to the substrate, we can recognize the cut-off point of the plating process, which prevented plating over the template. To fabricate the chassis, a composite consisting of gelatin with dispersed magnetic nanoparticles was molded into the remaining microchannels B. After curing the gelatin overnight, the entire structure was released by dissolving the photoresist layer in acetone. Optical microscope images (Figure 2B) verified the successful fabrication of the microrobot, which comprises the magnetic gelatin composite chassis interlocked with six electroplated iron helical micromachines. Scanning Electron Microscope (SEM) and Energy-Dispersive X-ray Spectroscopy (EDX) analyses further verified the composition of the superstructure (Figure 2D).

## 2.2. Pure gelatin chassis stability and solubility

The concentration of gelatin in the chassis significantly influenced its mechanical stability. Lower concentrations of gelatin ($\leq 5\%$) yielded microstructures with poor mechanical stiffness, which tended to collapse likely due to electrostatic interactions between the helical plated parts (Figure 2E, Fig. S5, Supporting Information). Conversely, structures with improved mechanical integrity were attained by increasing the gelatin concentration in the casting solution. Based on these experiments, we concluded that casting formulations with a gelatin content of 10% are adequate to manufacture microstructures with the required mechanical stability. Interestingly, pure gelatin cast structures exhibited a high resistance to solubilization in water even at elevated temperatures of more than 70 °C (Fig. S6, Supporting Information). As reported earlier, gelatin grown in confinement can display distinct mechanical properties and solubility characteristics in comparison to its bulk counterparts.[31] In bulk gelatin, the microscopic secondary structure consists of a combination of triple helices and random coils of collagen. However, when the gelation occurs in microsized spaces, random coils are less favored and turn into assembled β-sheet structures, which stabilize the hydrogel network, allowing it to preserve its solid-like features even at elevated temperatures.[32] However, when gelatin was mixed with magnetic nanoparticles, gelatin structures could dissolve upon heating (see further information in section 2.3).

## 2.3. Synthesis of optimized magnetic nanoparticles for gelatin composite dissolution

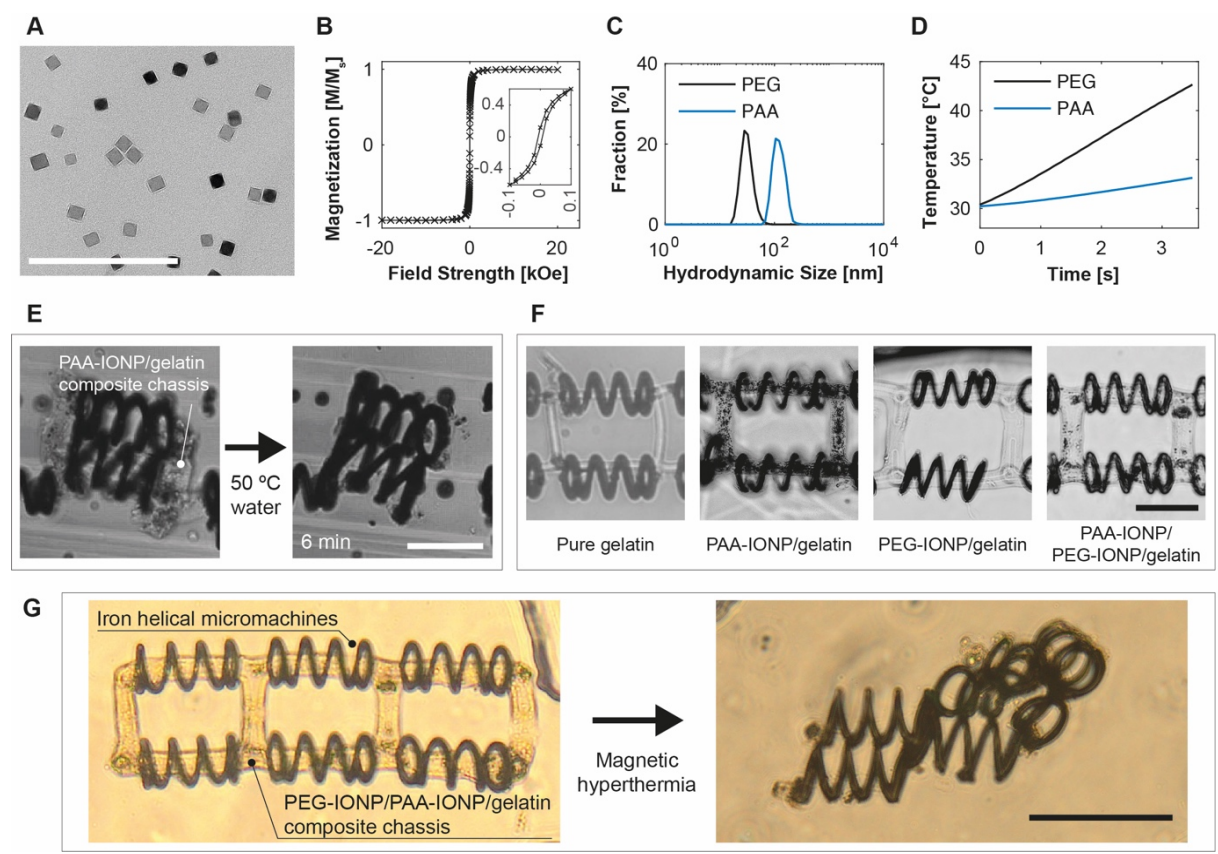

**Figure 3.** Characterization of the magnetic hyperthermia nanoparticles (A-D) and dissolution performance of their composites in gelatin matrix (E-G). (A) Transmission electron microscopy (TEM) images of the synthesized IONPs. The scale bar is 200 nm. (B) Magnetization hysteresis loop of IONPs. (C) Graph showing the direct light scattering (DLS) spectra of PAA-IONPs and PEG-IONPs. (D) Temperature change over time by applying high-frequency magnetic fields to PAA-IONPs and PEG-IONPs in water. (E) Micrographs illustrating the dissolution of PAA-IONP/gelatin composite chassis in 50 ºC water. The scale bar is 50 μm. (F) Interlocked structures with chassis containing different types of particles. From left to right: pure gelatin chassis, gelatin with PAA-IONPs, gelatin with PEG-IONPs and gelatin with a mixture of PAA- and PEG-IONP. Note that PEG-IONPs distribute very homogenously in gelatin, and displays similar finish with that of pure gelatin. Conversely, PAA-IONPs form significant agglomerates in gelatin. The scale bar is 50 μm. (G) Microscope images showing the dissolution of a chassis made of PEG-IONP/PAA-IONP/gelatin composite via magnetic hyperthermia (PAA-IONPs and PEG-IONPs were prepared in a 2:1 weight ratio.).

To develop a chassis capable of solubilization through magnetic hyperthermia and releasing its mechanically interlocked iron helical micromachines, we incorporated magnetic nanoparticles within the gelatin matrix to serve as remotely actuated heat transducers. The magnetic IONPs were synthesized through the thermal decomposition of tris(acetylacetonato)iron(III) using a protocol developed by Sun et al.[33] (see details of the synthetic procedure in the experimental section). **Figure 3**A displays a transmission electron microscopy (TEM) image of the synthesized particles, showing a highly uniform cubic morphology characterized by an average edge length of 17.2 ± 1.1 nm and a height of 23.3 ± 1.3 nm, respectively (Fig. S7a, Supporting Information). With a highly crystalline inverse spinel structure (XRD analysis in Fig. S7b, Supporting Information), these magnetic nanoparticles exhibited soft ferromagnetic behavior, with a coercivity of 6.72 Oe, as shown in the magnetic hysteresis loop (Figure 3B).

The synthesized IONPs were initially coated with oleic acid, making them hydrophobic. To ensure efficient dispersion of these hydrophobic nanoparticles in the gelatin matrix and improve their interaction with gelatin, we performed a subsequent ligand exchange process, by replacing the existing oleic acid with either PAA or PEG (FTIR and zeta potential analysis in Fig. S7c and d, Supporting Information). Direct light scattering (DLS) analysis confirmed that PEG-IONPs and PAA-IONPs had hydrodynamic diameters of 32 nm and 124 nm, respectively, thereby ensuring effective nanoparticle dispersion in de-ionized (DI) water (Figure 3C). Finally, we evaluated the magnetic hyperthermia capabilities of these nanoparticles under an alternating magnetic field at 510 kHz and 20 mT (Figure 3D). While PAA-IONPs exhibit 0.99 ºC s$^{-1}$ maximum temperature increase speed, PEG-IONPs display 3.5 times better performance (3.49 ºC s$^{-1}$) at a concentration of 73 mg/ml. We also conducted a biocompatibility test to evaluate the potential of these nanoparticles for biomedical applications. PEG-IONPs displayed high cell viability at a wide range of concentration, while PAA-IONPs showed decreased cell viability as particle concentration increased (Fig. S8, Supporting Information).

As previously reported, pure gelatin structures exhibited remarkable resistance to solubilization at elevated temperatures. However, we wanted to observe if this behavior was also observed when gelatin was casted together with magnetic nanoparticles. To this end, we casted PEG-IONPs/gelatin composites, and the interlocked structures were immersed in a water container and subjected to heating on a hotplate set at 50 °C. We observed that these structures remained insoluble in water (Fig. S9, Supporting Information). Conversely, PAA-IONP/gelatin composites dissolved upon heating (Figure 3E, Fig. S10, Supporting Information). We assume that the β-sheet structures become disrupted by physico-chemical changes in the gelatin solution (i.e., pH)[34] or the physical interactions between the nanoparticles' ligands and the collagen fibers.[35,36] These factors could influence the cross-linking between polymer chains, affect the protein chain mobility, and alter the secondary protein structure conformation. We also noted that the composites with PAA-IONPs displayed a significant number of agglomerates. In contrast, PEG-IONPs distributed homogeneously in gelatin (Figure 3F, Fig. S11, Supporting Information), which could marginally impact the β-sheet arrangements of the collagen fibers,[32] thus preventing the solubility of the composite, akin to pure gelatin.

Based on these results, we proceeded to evaluate the composite solubility through magnetic hyperthermia. Interestingly, PAA-IONP/gelatin composites could not be dissolved when applying high-frequency magnetic fields (Fig. S12, Supporting Information). As shown in figure 3D, PAA-IONPs do not exhibit optimal magnetic hyperthermia features unlike PEG-IONPs. To solve this issue, we prepared a gelatin composite containing a mixture of particles with the two ligands (PAA-IONPs:PEG-IONPs = 2:1), which enabled the chassis to solubilize upon magnetic hyperthermia within 1 hour (Figure 3G). Thus, we concluded that the chassis should comprise PAA-IONP/PEG-IONP/gelatin composite. Table 1 summarizes all these results for reader guidance.

**Table 1.** Effect of adding IONPs on dissolution of the gelatin chassis (O: dissolved, X: not dissolved).

| Chassis compound | 50 ºC water | Magnetic hyperthermia |
|---|---|---|
| Pure gelatin | X | X |
| PAA-IONP/gelatin composite | O | X |
| PEG-IONP/gelatin composite | X | -[a] |
| PAA-IONP/PEG-IONP/gelatin composite (PAA-IONPs:PEG-IONPs=2:1) | O | O |

[a] As dissolution was not observed in 50 ºC water, the magnetic hyperthermia test was not conducted.

## 2.4. Navigation and actuation of the microrobotic superstructures

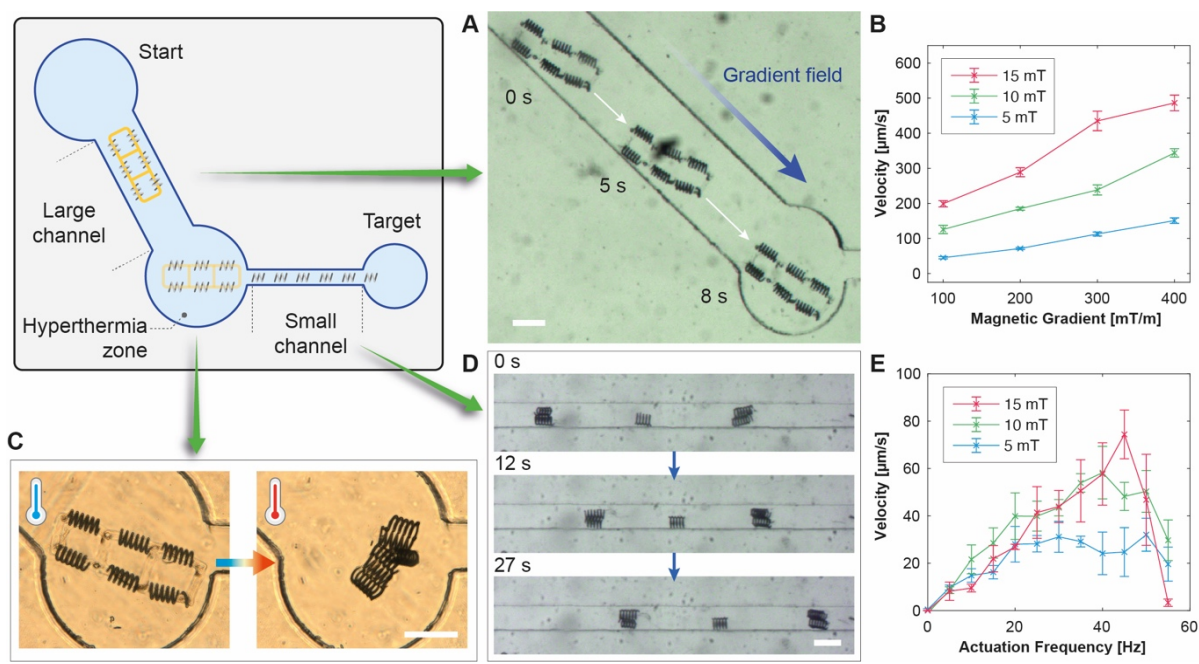

**Figure 4.** Demonstration of targeted delivery using a microrobotic superstructure through the depicted microfluidic tapered channel. (A) Optical microscope images of the microfluidic channel (width: 400 μm) employed for the navigation experiments of six helical micromachines and gelatin chassis powered by gradient magnetic field. (B) Graphs illustrating the velocity change of the robot (six helical micromachines and gelatin chassis) as a function of gradient field. Error bars represent the standard error. (C) Micrographs illustrating the on-command gelatin chassis dissolution by IONPs magnetic hyperthermia. (D) Optical microscope images of the small channel (width: 85 μm) design used for the navigation of six helical micromachines by rotating magnetic field. (E) Graph showing the velocity change of a helical micromachine as a function of actuation frequency. Error bars represent the standard error. Scale bars indicate 100 μm.

The locomotion of the interlocked microrobotic superstructures was investigated within a microfluidic chip with a channel with two sections of different widths, 400 μm and 85 μm. Application of a magnetic field gradient actuates the superstructures along the 400-μm section (**Figure 4**A, Movie S1, Supporting Information). Within the assessed magnetic field gradients and magnetic field strengths, the superstructure achieved up to 486 μm s$^{-1}$ of velocity at 400 mT m$^{-1}$ in a constant field of 15 mT (Figure 4B). Also, it was observed that the microrobot's velocity was directly proportional to the applied magnetic field gradient, which can be attributed to the dependency of the magnetization with the applied magnetic field. Within the larger section of the microchannel, the magnetic gradients served to navigate the superstructure into a target circular pocket, as shown in Figure 4C. The robot's progression was impeded at this point due to the decreased width of the subsequent channel section. At this point, the superstructure was subjected to high-frequency magnetic fields of 20 mT at 510 kHz for approximately an hour to dissolve the chassis (Figure 4C).

After the release, the helical micromachines were maneuvered using rotating magnetic fields, as illustrated in Figure 4D. While some helical micromachines exhibited adhesion among themselves, individual propulsion speeds were characterized as a function of the rotational field strength and frequency (Figure 4E, Movie S2, Supporting Information). At 15 mT, a clear step-out frequency was observed at 45 Hz, beyond which the forward velocity decreased drastically. A similar step-out frequency was observed at 40 Hz when actuated at 10 mT. However, we could not observe an obvious step-out frequency for helical micromachines actuated at 5 mT. Although the application of a gradient field generates faster motion (Fig. S13, Supporting Information) compared to using a corkscrew motion, this locomotion allows for more precise control in confined spaces. Furthermore, the corkscrew motion is a more efficient strategy when the size of the micromachine becomes smaller or the medium is denser.[37] Interestingly, the helical micromachines which were assembled exhibited actuation and forward propulsion (Movie S3, Supporting Information). This effect could be avoided by further treating the surface of the helical structures with additional coatings to promote repulsion. Nevertheless, it can result into an interesting feature that could be essentially engineered with different shapes to increase the magnetic volume of two released components by programmable re-assembly.

Finally, implementing the proposed delivery strategy of interlocking the multiple helical micromachines into a superstructure could significantly improve targeted delivery. Employing a superstructure as a carrier of helical machines enhanced the velocity a maximum of 4 times in contrast to using a single helical micromachine in the gradient field (Fig. S13, Supporting Information).

## 3. Conclusion

In this study, we demonstrated an efficient microscale targeted delivery approach using microrobotic superstructures that adapt their size to navigate through channels of varying dimensions. These superstructures were fabricated by interlocking electroplated-iron helical micromachines with a microcasted thermally responsive magnetopolymer chassis. Notably, this method enables batch fabrication of superstructures without additional assembly steps. The chassis, composed of thermally responsive gelatin infused with magnetic IONPs, which function as heat transducers, allowing for the controlled dissolution of the gelatin matrix on-command by means of high-frequency magnetic fields. To this end, we have used a combination of PAA-IONPs and PEG-IONPs to reach a tradeoff between nanoparticle dispersion within the gelatin matrix, while enhancing the thermal gelatin dissolution. By remotely instructing the dissolution of the gelatin, we enable the release of nested microhelices, facilitating navigation through narrower vessels.

The presented manufacturing method can be extended to other platable magnetic materials and other polymers or composites containing various functional nanoparticles or contrast agents, which could be of interest in the biomedical domain. We envision the adoption of this approach for the realization of more complex superstructures capable of performing intricate delivery procedures within the human body. Examples include accessing small capillaries, ducts in various organs, aneurysms, or the subarachnoid space, among others.

## 4. Methods

*Materials:* Iron(II)-sulfate heptahydrate ($FeSO_4 \cdot 7H_2O$), iron(II) chloride tetrahydrate ($FeCl_2 \cdot 4H_2O$), ammonium chloride ($NH_4Cl$), sulfuric acid ($H_2SO_4$), acetone, isopropyl alcohol and gelatin from porcine skin (gel strength 300, type A) were purchased from Sigma-Aldrich. AZ IPS 6090 photoresist and AZ 726 MIF developer were purchased from Microchemicals GmbH. Borosilicate substrate glass slides and Immersol 518f (Carl Zeiss Microscopy GmbH) were purchased from VWR International. Conductive Copper tape was purchased from Distrelec. All the other chemicals were purchased from Sigma Aldrich.

*Template Fabrication:* For the preparation of the template a recent report was used as a basis. As substrate for all the subsequent fabrication steps a borosilicate glass slide (30 mm diameter) was used. A conductive substrate was deposited on the glass a 20 nm titanium layer was sputtered onto the glass slide for adhesion before depositing an 80 nm Cu layer onto the titanium to enable the electrodeposition onto the glass substrate. Next, a positive tone AZ IPS 6090 photoresist was spin coated on the copper side of the glass slide for 17 seconds at 2000 rpm s$^{-1}$ with an acceleration of 300 rpm s$^{-1}$ to ensure a resist thickness of 45 μm. The coated substrate was baked on a hotplate at 125 °C for 5 minutes by ramping the temperature from room temperature at 8 °C per minute to evaporate the solvent. A commercial two-photon lithography system (Nanoscribe Photonic Professional GT, Nanoscribe GmbH) was used to write on the photoresist the microchannels. The design of the microstructure was imprinted into the photoresist by using a 63× objective (Carl Zeiss AG) and a refractive index matching oil (Immersol 518f, Carl Zeiss Microscopy GmbH). The design of the microstructure was processed by the DeScribe software (Nanoscribe GmbH) and the slicing / hatching distances were set to 0.3 μm. The laser power was set to 23% of the maximum laser power of the laser and the scan speed was set to 7500 μm s$^{-1}$. The strong reflection of the underlaying copper layer enabled the detection of the copper/photoresist interface, which is crucial for further fabrication steps. The microstructure had two types of channels, A and B. Channel A let the electrolyte reach the copper substrate leading to electrical contact, while channel B did not connect to the substrate. This setup allowed selective plating in channel A and filling channel B with gelatin. After light exposure and baking at 110 °C for 1 minute. Last, the substrate was developed in AZ MIF 726 to remove the exposed regions for 13 minutes to form the channels (Fig. S14, Supporting Information). Then, the template was immersed in water to remove the developer. The channels of the micro-mold were kept in water to keep the channels wet.

*Iron Deposition:* An electrolyte consisting of 250 g L$^{-1}$ of $FeSO_4 \cdot 7H_2O$, 42 g L$^{-1}$ of $FeCl_2 \cdot 4H_2O$ and 20 g L$^{-1}$ of $NH_4Cl$ mixed with 200 ml of DI water was used. The solution was thoroughly stirred at 250 rpm until the iron salts were dissolved. The pH value of the electrolyte was adjusted to a pH of 2 by adding sulfuric acid and the electrolyte was heated to 50 °C for iron deposition. A platinum sheet was used as an inert counter electrode and a double junction Ag/AgCl electrode was used as a reference electrode (Fig. S15, Supporting Information). The substrate was quickly immersed into the electrolyte to prevent the fast drying of the microchannels, ensuring a complete infiltration of the electrolyte within the channels and good electrical contact with the conductive copper substrate. The iron was electrodeposited potentiostatically by means of a potentiostat/galvanostat (PGSTAT204, Metrohm AG) by applying a constant potential of -0.95 V. Typical potentiostatic curves for this process show an initial sharp spike of current (capacitive charging of the double layer), which is followed by a steady current step. When the deposit reaches the exit of the microchannel, a sharp decrease of the current is detected. At this moment, the process was stopped, and the sample was removed and thoroughly rinsed in water to avoid the degradation of the plated iron structures caused by the remaining acidic electrolyte. After rinsing, the sample was set to dry at room temperature.

At this stage, the A microchannel was selectively filled with metallic iron and the microchannels, which were not connected to the substrate, remained unfilled.

*Nanoparticle Synthesis:* Magnetic iron oxide particles were synthesized according to a previous recipe with minor adjustments. In a typical synthesis, 530 mg iron(III) acetylacetonate (Fe(acac)$_3$), 199.6 mg sodium oleate and 1.664 g oleic acid were dispersed in a mixture of benzyl ether (7 ml), 1-octadecene (15 ml) and tetradecene (3 ml). The dispersion was degassed at 60 °C under vigorous stirring and then heated under constant N$_2$ flow by ramping the temperature at 3 °C min$^{-1}$ until reaching reflux at 294 °C for 90 minutes. After the reaction was completed, the slurry was cooled at room temperature. To remove organic impurities from the as-synthesized particles, the crude product was subsequently washed twice with a mixture of chloroform (25 ml) and acetone (75 ml). Subsequent purification was conducted by redispersing the particles in chloroform followed by precipitation in a mixture of methanol (50 ml) and acetone (50 ml) twice.

*Ligand exchange of nanoparticles with polyacrylic acid:* We conducted a ligand exchange procedure to substitute the oleic acid (hydrophobic) from the surface of the as-synthesized particles with polyacrylic acid (hydrophilic) as reported in previous protocols.[38] Briefly, 80 mg of synthesized particles were dispersed in 8 ml of THF (tetrahydrofuran). This dispersion was added dropwise into 72 ml of dry THF containing 2.88 g of PAA under vigorous stirring for 72 hours under a N$_2$ blanket to enable a complete ligan exchange. Afterwards, the particles functionalized with polyacrylic acid were collected by centrifugation and washed 5 times by redispersion and subsequent centrifugation in DI water. Finally, particles were dispersed in 20 ml DI water, deprotonated by the addition of 5 μl of NaOH (0.5 M), washed again by centrifugation, and redispersed in DI water at the desired concentration.

*Ligand exchange of nanoparticles with polyethylene glycol:* Within a N$_2$ filled glovebox 500 mg of mPEG-phosphate was dissolved in 72 ml of dry toluene inside a Teflon reactor. Subsequently 80 mg of oleic acid covered particles dispersed in 8 ml of toluene were added, before the addition of 8 ml of dry Ethanol under constant stirring. Finally, the Teflon reactor was sealed, sonicated for 10 minutes within an ultrasonic bath and the mixture was allowed to react overnight at 90 °C. After reaction the PEG functionalized particles were collected via extraction with DI-water and washed 5 times by centrifugation and redispersed in DI-water.

*Nanoparticle–Gelatin Infiltration:* For the nanoparticle-gelatin composites, 25 μl of PEG-coated nanoparticles were pipetted into an Eppendorf tube. The aqueous nanoparticle solution was centrifuged at 14500 rpm for 10 minutes and the supernatant was discarded. 25 μl of PAA-IONPs were added and mixed with the PEG-IONPs resulting in a 25 μl nanoparticle solution with 7.3 mg ml$^{-1}$ PAA-IONPs and 3.6 mg ml$^{-1}$ PEG coated nanoparticles, leading to an aqueous solution with 10.9 mg ml$^{-1}$ nanoparticles. The nanoparticles were redispersed in a sonicator for 10 minutes and heated up to 50 °C before adding 2.5 mg of type A gelatin from porcine skin. The nanoparticle-gelatin mixture was sonicated for an hour at 50 °C before pipetting the solution onto the template. For the infiltration of the nanoparticle-gelatin solution, 5 cycles of 0.05 MPa vacuum at 65 °C were applied to ensure successful infiltration without the evaporation of the aqueous part of the solution. After the infiltration, the sample was stored at 5 °C over night.

*Magnetometry:* The magnetic characteristics of the particles were assessed using vibrating sample magnetometer (VSM) measurements on powder samples. The measurement was conducted in a magnetic field range of up to 40 kOe. Hysteresis loop measurements were conducted at room temperature with a maximum field strength of 20 kOe.

*Transmission electron microscopy (TEM):* TEM images were captured using an FEI Talos F200X (Chem S/TEM) operating at 200 kV, equipped with an X-FEG emitter and CETA camera (16M pixel CMOS Camera). Specimens for imaging were prepared by dispersing particles in a diluted solution of deionized water and depositing them onto a carbon-coated 400 mesh TEM grid. The average diameter and standard deviation of particles were determined by measuring a minimum of 50 particles.

*X-ray Diffraction (XRD):* The crystalline structure of the samples was analyzed using X-ray diffraction with an Empyrean instrument (Malvern Panalytical) operating at room temperature. The X-ray source utilized copper radiation ($\lambda = 1.5406$ Å), and a PIXcel detector was employed. Sample preparation involved drop-casting a particle dispersion in ethanol onto a silicon holder.

*Dynamic Light Scattering and ζ-potential:* The particle's hydrodynamic diameter and ζ-potential were determined using a Zetasizer Nano-ZS.

*Fourier-Transform Infrared Spectroscopy (FTIR):* FTIR spectra measured using a Bruker Tensor 27 spectrometer across the range of 4000-400 cm$^{-1}$ with a resolution of 4 cm$^{-1}$.

*Thermogravimetric Analysis (TGA):* Thermogravimetric measurements were conducted using a Mettler Toledo TGA/DSC 3+ Star System. Measurements were taken at temperatures of up to 1173 K under constant airflow.

*Stripping:* To release the interlocked structures from the substrate, the photoresist template was first stripped by placing it in a specific holder at a 45° angle, which was carefully immersed in a glass beaker filled with acetone for 5 minutes. After the photoresist was clearly dissolved, the sample holder was carefully transferred to a glass beaker filled with DI water. After 2 minutes, the sample was carefully extracted from the glass beaker and was quickly wetted with DI water with a pipette. At this point, the sample was always kept wet to avoid the dehydration of the gelatin. The sample was transferred into a petri dish filled with DI water. A PDMS microfluidic chip was immersed in the same petri dish and placed next to the sample. The interlock structures were carefully detached from the substrate and were transferred to the PDMS microfluidic chip using a microprobe while always being fully immersed in DI water. Once the interlocked structure was released into the microfluidic chip, the chip was transferred to another petri dish. The microfluidic chip remained immersed in water throughout all the experiments.

*Scanning Electron Microscopy:* Scanning electron microscope images were obtained using a Zeiss ULTRA 66 at 5 kV with a 30 aperture and EDX elemental mapping was done at 20 kV, 60 μm aperture and 10 mm working distance.

*Microfluidic Chip Fabrication:* The microfluidic chips were fabricated using photolithography and PDMS casting. A film mask was prepared containing the design of a chip with the length of 16.5 mm and channel widths ranging from 80 μm to 400 μm. 250 μm of SU-8 100 (Kayaku Advanced Materials) photoresist was used as a mold. Then, a 1-mm thick layer of PDMS was cast and subsequently cured at 80 °C for 2.5 hours. Additionally, a microscopic glass slide was also coated with a layer of PDMS and cured with the same curing parameters. Next, the PDMS chip was released from the silicon wafer. Additionally, the PDMS coated microscopic glass was cut to the size of the PDMS chip. The PDMS-coated glass slide and the PDMS chip were then plasma bonded and stored at 50 °C over night. After bonding, the entry channels were punched into the chip allowing the transfer of the superstructures into the chip. The use of a PDMS-coated substrate decreased the sticking of the gelatin to the chip.

*Actuation:* Electromagnetic systems (MFG-100 and MFG-100-I, Magnebotix AG) were used to magnetically actuate the microrobots. For the actuation of the iron helices rotating magnetic fields and for the actuation of the interlocked superstructures, magnetic gradients were generated with field strengths between 5 mT and 15 mT. Prior to actuating the microrobots, they were set free by stripping the photoresist and the sample was directly transferred into DI water. The microrobots were then transferred into a PDMS microfluidic chip with the use of a micromanipulator stage. The microfluidic chip was then placed in the working space of the magnetic coil setups and the magnetic actuation was characterized through the optical magnifiers.

*Magnetic hyperthermia:* To externally release the helical micromachines from the gelatin chassis, a magnetic hyperthermia setup was used. The hyperthermia setup (NAN201003 Magnetherm, nanoTherics Ltd.) was placed in an incubator (Galaxy 170 S, New Brunswick) to thermostatize the coil of the system at 37 °C during the heating experiments and ensuring that heating on the specimens was only due to hyperthermia generated by the nanoparticles. The capacitor mounted was 22 nF, a resonant frequency of 510 kHz was set on the function generator and the voltage was ramped up to 20 V with a power supply. Prior to setting the output, the PDMS microfluidic chip containing the interlocked structure was placed in a petri dish filled with DI water preheated to 37 °C. The petri dish containing the interlocked structure was placed in the center of the coil to get maximum magnetic field exposure. After switching on the output, the current in the coils increased to around 14 A.

**Acknowledgements**

This project has received funding from the European Union's Horizon 2020 Proactive Open program under FETPROACT-EIC-05-2019 ANGIE (No. 952152), the Swiss National Science Foundation under project number 198643, and ETH under grant number 22-2 ETH-040. M. K. acknowledges partial financial support by the Swiss National Science Foundation under project No. 200021L_197017. J.P-L. acknowledges the Agencia Estatal de Investigación (AEI) for the María de Maeztu, project no. CEX2021-001202-M and the grant PID2020-116612RB-C33 funded by MCIN/ AEI /10.13039/501100011033. J.L.-W. acknowledges his PhD grants in the framework of the project BeMAGIC funded by the European Union's Horizon 2020 Research and Innovation Programme under the Marie Skłodowska-Curie grant agreement No. 861145.

The authors would also like to thank the Scientific Center for Optical and Electron Microscopy (ScopeM) and the FIRST laboratory at ETH for their technical support, and the Cleanroom Operations Team of the Binning and Rohrer Nanotechnology Center (BRNC) for their help and support. Further Prof. Herman of the Nanoparticle Systems Engineering Lab and Prof. Pratsinis of the Particle Technology Laboratory are acknowledged for the access to various characterization equipment.